\documentclass[prl,twocolumn,showpacs,aps]{revtex4-1}

\usepackage{amsmath, amsfonts, amssymb, mathrsfs}
\usepackage{txfonts}
\usepackage{graphicx}
\usepackage{epsfig}
\usepackage{enumerate}
\usepackage{caption}
\usepackage{subcaption}

\usepackage{slashed}
\usepackage{ulem}

\newcommand{\bea}{\begin{eqnarray}}
\newcommand{\eea}{\end{eqnarray}}

\newcommand{\be}{\begin{equation}}
\newcommand{\ee}{\end{equation}}

\usepackage{ifpdf}
\ifpdf
\usepackage{epstopdf}
\fi

\usepackage[unicode]{hyperref}
\hypersetup{
	pdftitle={Quantum-Corrected Holographic Wilson Loop Expectation Values and Super-Yang-Mills Confinement},
	pdfauthor={Liu, Nian, Yue, Zheng},
	pdfsubject={Quantum-Corrected Holographic Wilson Loop Expectation Values and Super-Yang-Mills Confinement},
	colorlinks=true,
	linkcolor=blue,
	citecolor=blue,
	filecolor=black,
	urlcolor=blue
}

\def\url#1{}

\makeatletter
\newcommand{\vast}{\bBigg@{4}}
\newcommand{\Vast}{\bBigg@{5}}
\makeatother

\usepackage{bm}
\usepackage{color}

\begin{document}

\title{Quantum-Corrected Holographic Wilson Loop Expectation Values\\ and Super-Yang-Mills Confinement}
\author{Xiao-Long Liu}
\author{Cong-Yuan Yue}
\author{Jun Nian}
\email{nianjun@ucas.ac.cn}
\author{Wenni Zheng}
\affiliation{International Centre for Theoretical Physics Asia-Pacific, University of Chinese Academy of Sciences, 100190 Beijing, China}

\begin{abstract}
Confinement is a well-known phenomenon in the infrared regime of (supersymmetric) Yang-Mills theory. While both experimental observations and numerical simulations have robustly confirmed its existence, the underlying physical mechanism remains elusive. Unraveling the theoretical origin of confinement continues to be a profound and longstanding challenge in both physics and mathematics. Motivated by recent advances in quantum Jackiw-Teitelboim gravity, we investigate the Wilson loop expectation values in the large-$N$ limit of $\mathcal{N}=4$ super-Yang-Mills theory at finite chemical potential, employing a holographic approach within the background of an extremal AdS$_5$ Reissner-Nordstr\"om black brane. Our results reveal that quantum gravitational fluctuations in the near-horizon region significantly modify the holographic Wilson loop expectation values. These values exhibit an area-law behavior, indicative of a confining quark-antiquark potential. Within this framework, our findings suggest that confinement in the super-Yang-Mills theory arises as a consequence of near-horizon quantum gravity fluctuations in the bulk extremal AdS$_5$ black brane geometry.
\end{abstract}

\maketitle

\textit{Introduction.---} Ever since its establishment \cite{Maldacena:1997re}, the AdS/CFT correspondence has served as an important tool in understanding the strong-coupling dynamics of the supersymmetric Yang-Mills (SYM) theory in the infrared (IR) regime. According to the AdS/CFT correspondence, a Wilson loop $\mathcal{C}$'s quantum expectation value in the large-$N$ limit of $\mathcal{N} = 4$ super-Yang-Mills theory can be computed holographically via \cite{Rey:1998ik, Maldacena:1998im}:
\be\label{eq:W from S_NG}
  \langle W (\mathcal{C}) \rangle \simeq e^{- S_{\text{\text{NG}}}}\, ,
\ee
where $S_{\text{NG}}$ is the Nambu-Goto (NG) action of the fundamental string worldsheet in the anti-de-Sitter spacetime anchored to the Wilson loop $\mathcal{C}$ on the boundary. Hence, the minimal surface provides the dominant contribution.

On the other hand, for a temporal rectangular Wilson loop in the fundamental representation with the time interval $T \to \infty$, $\langle W (\mathcal{C}) \rangle$ obeys the following relation:
\be\label{eq:W and Vqqbar}
  \langle W (\mathcal{C}) \rangle \simeq e^{-T\cdot V_{q \bar{q}} (L)}\, ,
\ee
where $V_{q \bar{q}} (L)$ is the potential between a heavy quark-antiquark pair, which can be extracted by combining \eqref{eq:W from S_NG} and \eqref{eq:W and Vqqbar}. If the potential $V_{q \bar{q}} (L)$ for long-distance $L$ behaves linearly in $L$, it is an indication of confinement in SYM theory.

The first attempts along this line were made in \cite{Rey:1998ik, Maldacena:1998im}, where the authors found that a strictly AdS spacetime leads only to a Coulomb potential without a confining linear term. Later, it was pointed out in \cite{Witten:1998zw, Rey:1998bq, Brandhuber:1998bs, Greensite:1999jw} that for a black hole with finite temperature, besides the Coulomb potential, the potential $V_{q \bar{q}} (L)$ can contain some confining terms, consistent with the well-known results in the literature \cite{Polyakov:1978vu, Susskind:1979up}. Similar results have been found and extended in many following works, e.g., \cite{Kinar:1999xu, Bigazzi:2004ze}. However, a holographic explanation for confinement at zero temperature remains open. Phenomenologically, we can modify the bulk metric of the AdS spacetime by hand. Some proposals of the modified AdS metrics successfully led to a confining potential \cite{Andreev:2006ct, Karch:2006pv, Andreev:2007vn, Pirner:2009gr, Nian:2009mw, Trawinski:2014msa}, but first-principle derivations and more fundamental understanding are still lacking.

The recent progress on nearly AdS$_2$ quantum gravity brings new ideas to this problem. As a toy model of quantum gravity, the Jackiw-Teitelboim (JT) gravity on AdS$_2$ is governed by the Schwarzian theory on the 1d boundary \cite{Almheiri:2014cka, Maldacena:2016upp}. This approach can be applied to the near-horizon AdS$_2$ region of a higher-dimensional near-extremal Reissner-Nordstr\"om (RN) AdS$_D$ ($D \geq 4$) black hole or black brane, which allows us to study certain quantum gravity fluctuations, i.e., the Schwarzian modes.

After quantum averaging these modes, the spacetime metric close to the boundary receives mild quantum corrections \cite{Blommaert:2019hjr}, and so does the holographic $\langle W \rangle$. Consequently, the quark-antiquark potential naturally acquires a confining linear potential at zero temperature, indicating the SYM theory's confinement in the IR regime. Hence, within the AdS/CFT correspondence framework, the linear confinement at zero temperature is due to the near-horizon quantum gravity effects. In the phase diagram of SYM theory, our new result can be interpreted as deforming the superconformal $\mathcal{N}=4$ SYM theory by finite chemical potential, in contrast to the finite-temperature deformation considered in \cite{Witten:1998zw, Rey:1998bq, Brandhuber:1998bs, Greensite:1999jw}.

\begin{figure}[htb!]
\begin{center}
\includegraphics[width=5.5cm, angle=0]{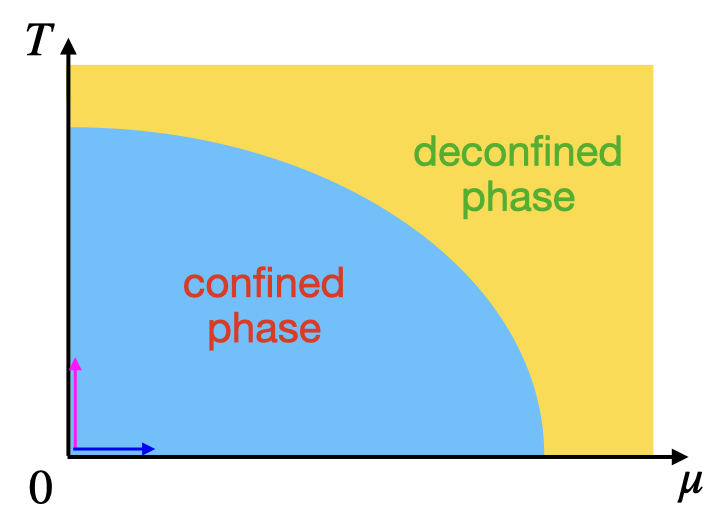}
\caption{The schematic phase diagram of SYM theory (see e.g. \cite{Yamada:2006rx, Evans:2010iy})}\label{fig:PhaseDiagram}
\end{center}
\end{figure}

\textit{Quantum-Corrected Spacetime from JT Gravity.---} We start with a near-extremal Reissner-Nordstr\"om AdS$_5$ black brane, whose near-horizon region is AdS$_2 \times \mathbb{T}^3$. To introduce quantum gravity fluctuations in the near-horizon AdS$_2$ region, we apply the approach introduced in \cite{Maldacena:2016upp, Blommaert:2019hjr} by treating the metric factor as a bilocal operator. Consequently, the quantum-averaged Euclidean AdS$_2$ metric is given by
\be
  \langle ds^2\rangle = \langle G_{\partial\partial}(2\zeta)\rangle_\beta \cdot L^2_2\Big(d\zeta^2 + dt_E^2 \Big)\, ,
\ee
where the quantum-averaged bilocal operator at the inverse temperature $\beta$ is
\begin{align}
  \langle G_{\partial\partial}(2 \zeta)\rangle_{\beta} & = \frac{1}{(2C)^2}\int^{\beta}_0 dM\, {\rm sinh} \Bigl(2\pi\sqrt{M}\Bigr)\, e^{-\beta M} \nonumber\\
  {} & \cdot\int^{\infty}_0 dE\, {\rm sinh} \Bigl(2\pi\sqrt{E}\Bigr)\, e^{-2 \zeta (E-M)}\, \Gamma\Bigl(1\pm i\sqrt{M}\pm i\sqrt{E}\Bigr)\, ,
\end{align}
and $\Gamma\bigl(1\pm ix\bigr) \equiv \Gamma\bigl(1+ix\bigr)\, \Gamma\bigl(1-ix\bigr)$. Taking the extremal limit $\beta \to \infty$ (implying $M \to 0$) and adopting the reparametrization $E = 2C (M+\omega)$ and $\zeta \to \zeta / (2C)$, we obtain for $\omega\gg M$:
\begin{align}
  \langle G_{\partial\partial}(2 \zeta)\rangle_{\infty} & = \frac{1}{2C}\int_0^\infty d\omega\, {\rm sinh} \Bigl(2\pi\sqrt{2C\omega}\Bigr)\, e^{-2 \zeta \omega}\, \Gamma(1\pm i\sqrt{2C\omega})^2.
\end{align}
Finally, the complete quantum-corrected Euclidean AdS$_2$ metric is
\begin{equation}
  ds^2 = \frac{h(\zeta)}{\zeta^2}\cdot L^2_2(d\zeta^2 + dt_E^2)\, ,
\end{equation}
with the factor $h(\zeta)$ given by
\begin{align}\label{eq:hz1}
  h(\zeta) & = -\frac{\zeta^2}{\pi^2 C} \int^{\infty}_0 d\omega\, {\rm sinh} \Bigl(2\pi \sqrt{2C\omega}\Bigr)\, e^{-2 \zeta \omega}\, \Gamma\Bigl(1\pm i\sqrt{2C\omega}\Bigr)^2,
\end{align}
which approaches the classical value $1$ towards the AdS$_2$ boundary $\zeta \to 0$. The expression \eqref{eq:hz1} can be viewed as a Fourier transform of a function $\widetilde{h} (\omega)$. Later in this paper, we focus on small quantum corrections that deviate from the classical background. Hence, we expand $\widetilde{h} (\omega)$ in the classical limit ($C^{-1} \ll \omega$) and then perform the Fourier transform. In this limit, the quantum correction factor $h(\zeta)$ becomes
\begin{equation}
h(\zeta) = 1 + \frac{3}{2 \pi^4} \left(\frac{\zeta}{C}\right)^2 - \frac{15 }{2 \pi ^6 }\left(\frac{\zeta}{C}\right)^3 + \frac{315}{8 \pi ^8 }\left(\frac{\zeta}{C}\right)^4 +\cdots\, .
\end{equation}

Now, we consider the quantum corrections to the AdS$_5$ black brane metric. The RN Euclidean AdS$_5$ black brane has the metric and the background gauge field:
\begin{align}
ds^2 & = \frac{u^2}{L_{\text{AdS}}^2} f(u)\, dt_E^2 + \frac{L_{\text{AdS}}^2}{u^2} \frac{du^2}{f(u)}+\frac{u^2}{L_{\text{AdS}}^2} d\vec{x}\,^2\, ,\label{eq:RN BH metric 1}\\
A & = \mu\left(1-\frac{u_T^2}{u^2}\right) dt\quad \text{with}\quad \mu=\frac{\sqrt{3}}{2}\frac{g_s\, u_T}{L_{\rm AdS}^2}\, Q\, ,
\end{align}
where $g_s$ is the dimensionless coupling constant of the gauge field, and $u_T$ denotes the horizon position. The factor $f(u)$, the mass $M$, and the temperature $T_h$ are
\be
f(u) = 1 - M\frac{u_T^4}{u^4} + Q^2\frac{u_T^6}{u^6},\quad M=1+Q^2,\quad T_h=\frac{(2-Q^2)u_T}{2\pi L_{\text{AdS}}^2}\, .
\ee
From now on, we focus on the extremal case ($T=0$) and set $Q^2=2$.

Next, we introduce quantum gravity fluctuations in the near-horizon AdS$_2$. By imposing continuity and smoothness conditions in the overlap region, we obtain the quantum-corrected extremal RN Euclidean AdS$_5$ metric:
\begin{equation}\label{eq:quantum AdS5 metric Euclidean}
  ds^2 = \alpha' \left[\frac{U^2}{R^2} \left(f(U)\, h(U)\, dt_E^2 + d\vec{x}\,^2 \right) + R^2 f^{-1}(U)\, h(U)\, \frac{dU^2}{U^2}\right]\, ,
\end{equation}
with the quantum correction factor
\begin{align}
h(U) & = 1 + \frac{R^4}{96\, \pi ^4 C^2 (U-U_T)^2} -\frac{5 R^6}{1152\, C^3 \left(\pi ^6 (U-U_T)^3\right)} \nonumber\\
{} & \quad + \frac{35 R^8}{18432\, \pi ^8 C^4 (U-U_T)^4}+\cdots\, ,
\end{align}
where $U \equiv u / \alpha'$, $U_T \equiv u_T / \alpha'$, and $R^2 \equiv L_{\text{AdS}}^2 / \alpha'$ with $\sqrt{\alpha'}$ denoting the effective string length. More details can be found in the supplemental materials.

\textit{Temporal Wilson Loop.---} We apply the same method as in \cite{Maldacena:1998im} to calculate the expectation value of a rectangular temporal Wilson loop, whose temporal and spatial sides have the lengths $T$ and $L$, respectively. We use the quantum-corrected AdS$_5$ metric \eqref{eq:quantum AdS5 metric Euclidean} and drop the subscript of $t_E$ for simplicity. This metric works well when the worldsheet is far from the horizon.

In the presence of an $U(1)$ gauge field, besides the Schwarzian modes, the gauge mode fluctuations should also provide quantum corrections \cite{Davison:2016ngz}. However, since the characteristic scale $M_{U(1)}$ of the gauge modes is much greater than the characteristic scale $M_{SL(2)}$ of the Schwarzian modes \cite{Iliesiu:2020qvm}, the $U(1)$ gauge fluctuations are effectively frozen in the low-temperature regime $T\ll M_{U(1)}$. In this paper, we only consider the quantum corrections from Schwarzian modes.

We choose the worldsheet coordinates $\tau=t$ and $\sigma=x$. The induced metric is
\begin{equation}
g_{\alpha\beta} = 
\begin{pmatrix}
g_{tt}(\partial_tt)^2 & 0\\
0 & g_{UU}(\partial_xU)^2 + g_{xx}(\partial_xx)^2\\
\end{pmatrix}\, ,\quad \alpha,\, \beta \in \{\tau,\, \sigma \}\, .
\end{equation}
Consequently, the NG action becomes
\be
  S_{\text{NG}} = \frac{T}{2\pi} \int_{-L/2}^{L/2} dx\sqrt{h^2(U)\, (\partial_x U)^2 + (U^4/R^4) f(U)\, h(U)}\, .
\ee
This action has the conserved charge
\begin{equation}
\frac{\frac{U^4}{R^4}f(U)\, h(U)}{\sqrt{(\partial_xU)^2 h^2(U)+\frac{U^4}{R^4}f(U)\, h(U)}} = \sqrt{\frac{U_0^4}{R^4}f(U_0)\, h(U_0)}\,.
\end{equation}
Thus, we have the relation between $x$ and $U$:
\begin{equation}\label{eq:x1}
  x = \frac{R^2}{U_0} \int_1^{U/U_0} \frac{h_y(y)\, dy}{y^2 \sqrt{\frac{f_y^2(y)\, h_y^2(y)}{f_y(1)\, h_y(1)} y^4 - f_y(y)\, h_y(y)}}\,,
\end{equation}
with
\begin{align}
  f_y(y) & = 1 - (1 + Q^2)\frac{U^4_T}{U^4_0y^4} + Q^2\frac{U^6_T}{U^6_0y^6}\, ,\\
  h_y(y) &= 1 + \frac{R^4}{96 \pi ^4 C^2 (yU_0-U_T)^2} -\frac{5 R^6}{1152 C^3 \left(\pi ^6 (yU_0-U_T)^3\right)} \nonumber\\
  {} & \quad + \frac{35 R^8}{18432 \pi ^8 C^4 (yU_0-U_T)^4}+\cdots\, ,
\end{align}
where $y \equiv U/U_0$, while $U_0$ and $U_T$ denote the extremum position of the string and the horizon position, respectively.

\begin{figure}[h]
    \centering
    \includegraphics[width=0.75\linewidth]{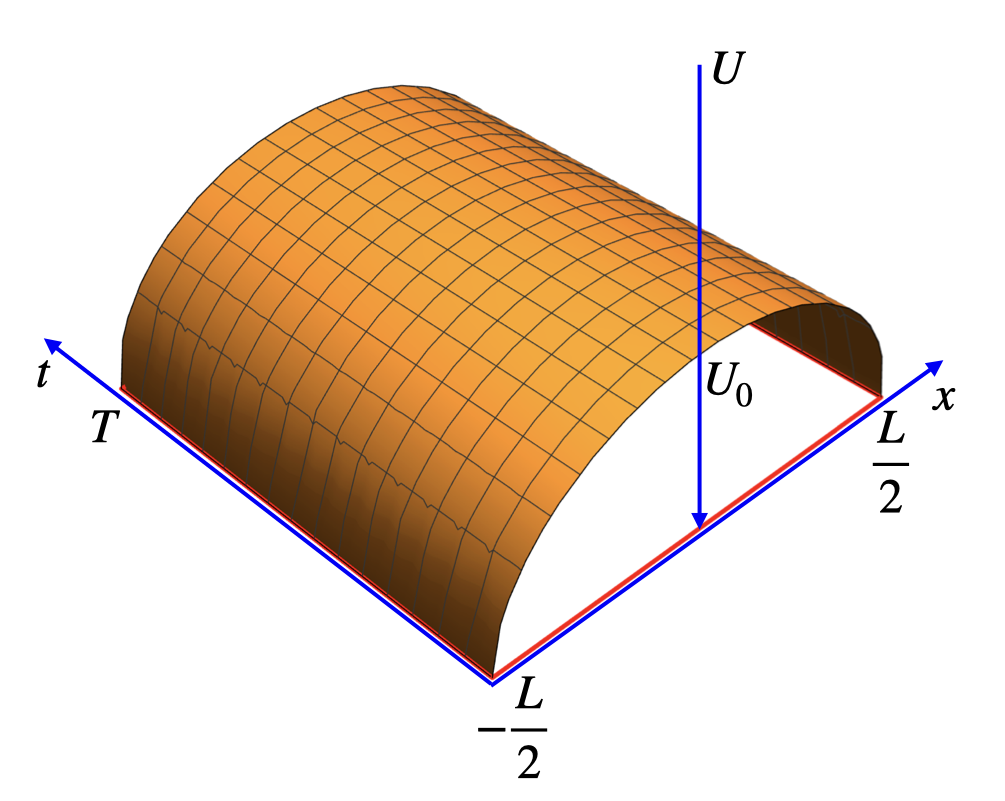}
    \caption{The minimal surface anchored to the rectangular temporal Wilson loop (denoted by the red line)}
    \label{RectangularWilsonLoop}
\end{figure}

As shown in Fig.~\ref{RectangularWilsonLoop}, once we fix the endpoints of the string on the boundary, the minimal surface in the bulk is also fixed. The distance $L$ between the quark and the antiquark obeys
\begin{equation}\label{eq:1loopsquarel}
\frac{L}{2} = \frac{R^2}{U_0} \int_1^{\infty} \frac{h_y(y)\, dy}{y^2\sqrt{\frac{f_y^2(y)\, h_y^2(y)}{f_y(1)\, h_y(1)} y^4 - f_y(y)\, h_y(y)}}\,.
\end{equation}
In order to evaluate this integral analytically, we expand the integrand for small $U_T/U_0$ and large $C$ as 
\begin{align}
\frac{L}{2} = \frac{R^2}{U_0} \int^{\infty}_1 dy\, \Bigg[ & \frac{1}{y^2 \sqrt{y^4-1}} + \frac{R^4 \left(y^4+y^2+1\right)}{192\, \pi ^4 C^2 U_0^2 y^4 \left(y^2+1\right) \sqrt{y^4-1}} \nonumber\\
  {} & + \mathcal{O} \left(C^{-3},U_T/U_0\right)\Bigg]\, ,
\end{align}
where the leading-order result is the same as in \cite{Maldacena:1998im}.

On the other hand, the static energy $E = S_{\text{NG}} / T$ is
\begin{align}
  E = \frac{2U_0}{2\pi} \Bigg[ & \int^{\infty}_1 dy \left(\frac{y^2 f_y(y)\, h_y^2(y)}{\sqrt{f_y^2(y)\, h_y^2(y)\, y^4 - f_y(y)\, h_y(y)\, f_y(1)\, h_y(1)}} - 1\right)\nonumber\\
  {} & - 1\Bigg]\, .\label{eq:1loopsquaree}
\end{align}
 We also expand the integrand for small $U_T/U_0$ and large $C$:
\begin{align}
E = \frac{U_0}{\pi} \Bigg[\int^{\infty}_1 dy\, & \Bigg(\frac{y^2}{\sqrt{y^4-1}} + \frac{R^4 \left(2 y^2+3\right)}{192\, \pi ^4 C^2 U_0^2 \left(y^2+1\right) \sqrt{y^4-1}}\nonumber\\
{} & - 1\Bigg) - 1\Bigg] + \mathcal{O} (C^{-3},U_T/U_0)\, .
\end{align}

Eq.~\eqref{eq:1loopsquarel} relates $L$ and $U_0$, while Eq.~\eqref{eq:1loopsquaree}  relates $E$ and $U_0$. In principle, we can combine these two equations to eliminate $U_0$ and obtain a relation between $E$ and $L$. A subtlety is that there are three different solutions when expressing $U_0$ in terms of $L$ via Eq.~\eqref{eq:1loopsquarel}. Fortunately, there is only one solution leading to the real result:
\begin{equation}\label{eq:Rectangular Wilson Loop Potential}
E = - \frac{4 \pi ^2}{\Gamma \left(\frac{1}{4}\right)^4}\frac{R^2}{L} + \frac{5\, \Gamma \left(\frac{1}{4}\right)^2}{4608\, \pi^5\, \Gamma \left(\frac{3}{4}\right)^2}\frac{R^2}{C^2}L +  \mathcal{O} \left(C^{-3}, \frac{U_T}{U_0}\right)\, ,
\end{equation}
which is just the desired quark-antiquark potential at zero temperature with both a Coulomb term and a linear term. The appearance of the confining linear term is due to the conformal symmetry breaking by the near-horizon quantum fluctuations, and the quantum scale $C^{-1}$ indicates its origin from the near-horizon quantum gravity fluctuations.

To double-check the analytic expression \eqref{eq:Rectangular Wilson Loop Potential}, we also evaluate Eq.~\eqref{eq:1loopsquarel} and Eq.~\eqref{eq:1loopsquaree} numerically, where the analytic expression (blue curve) and the numerical result (red dots) clearly show linear behavior for large $L$, significantly differing from the Coulomb potential (green line).
\begin{figure}[htb!]
\begin{center}
\includegraphics[width=7cm, angle=0]{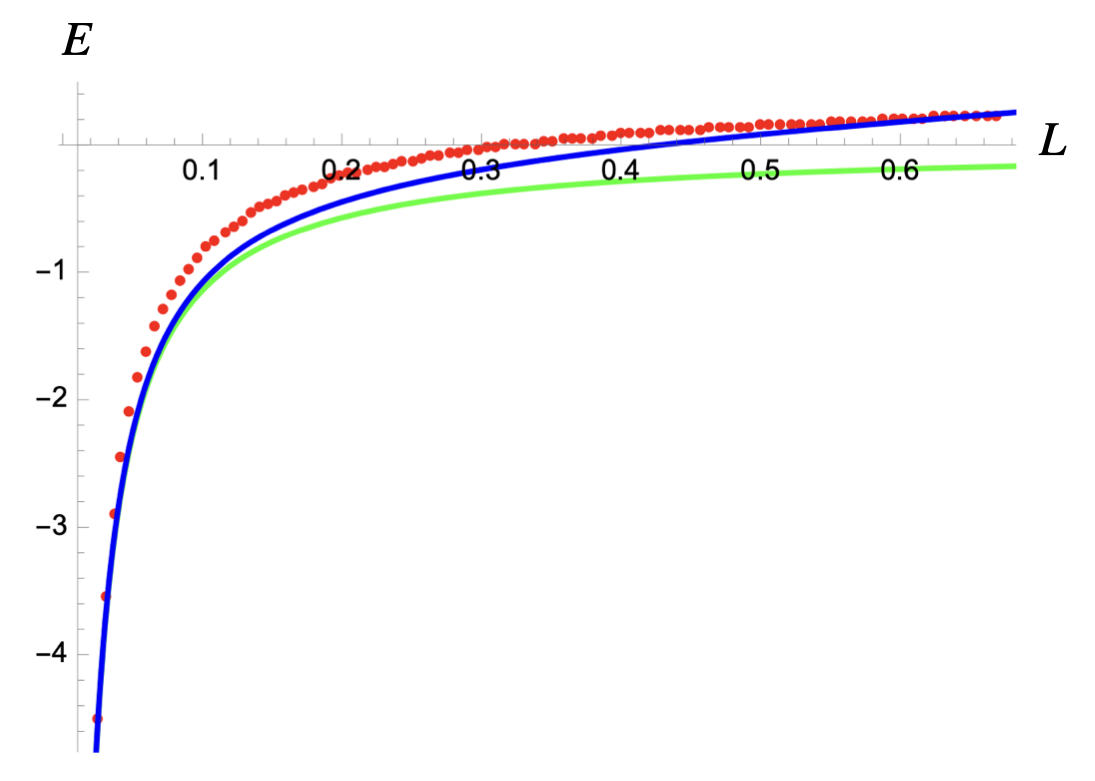}
\caption{The numerical result of the quark-antiquark potential from the rectangular temporal Wilson loop with the parameters $R=1$, $Q=\sqrt{2}$, and $C=0.01$}\label{fig:square}
\end{center}
\end{figure}

\begin{figure}[htb!]
    \centering
    \includegraphics[width=0.47\linewidth]{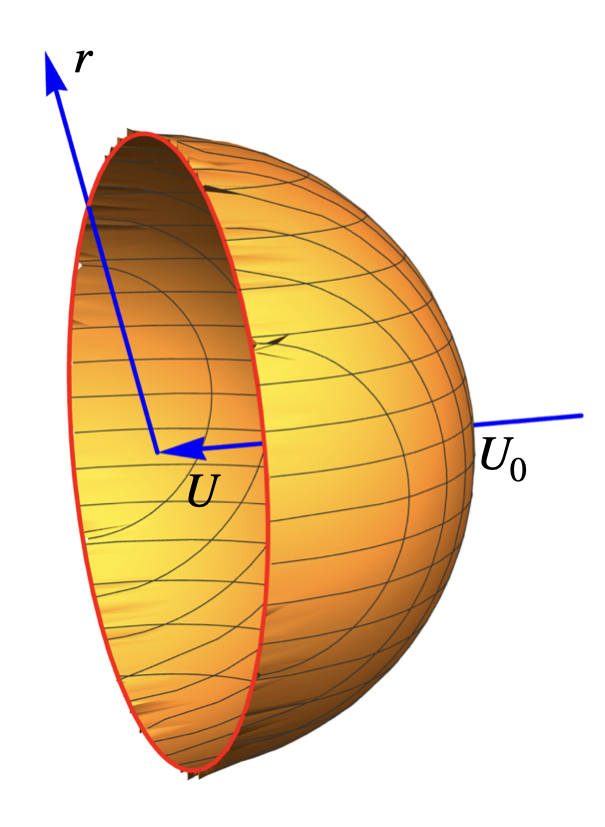}
    \caption{The minimal surface anchored to the circular spatial Wilson loop (denoted by the red line)}
    \label{CircularWilsonLoop}
\end{figure}

\textit{Spatial Wilson Loop.---} Now, let us consider a circular spatial Wilson loop with radius $L$ (Fig.~\ref{CircularWilsonLoop}). The quantum-corrected Euclidean AdS$_5$ metric with cylindrical coordinates $(r, \phi, x)$ is
\begin{align}
  ds^2 = \alpha' \Bigg[ & \frac{U^2}{R^2} \Big(f(U)\, h(U)\, dt^2 + dx^2 + dr^2 + r^2\, d\phi^2\Big)\nonumber\\
  {} & + R^2\, f^{-1}(U)\, h(U) \frac{dU^2}{U^2}\Bigg]\, .
\end{align}

We choose the worldsheet coordinates $\tau=\phi$ and $\sigma=r$. The induced metric is
\begin{equation}
g_{\alpha\beta} = 
\begin{pmatrix}
g_{\phi\phi}(\partial_{\phi}\phi)^2 & 0\\
0 & g_{UU}(\partial_rU)^2 + g_{rr}(\partial_rr)^2\\
\end{pmatrix}\, ,\quad \alpha,\, \beta \in \{\tau,\, \sigma \}\, .
\end{equation}
The NG action is then
\be
 S_{\text{NG}} = \int_0^L r dr\sqrt{(\partial_rU)^2 f^{-1}(U)\, h(U)+U^4/R^4}\, ,\label{eq:circular1loope}
\ee
which has the conserved charge
\begin{equation}
\frac{\frac{U^4}{R^4}}{\sqrt{(\partial_rU)^2f^{-1}(U)h(U)+\frac{U^4}{R^4}}} = \frac{U^2_0}{R^2}\, .
\end{equation}
This expression with $y \equiv U / U_0$ leads to
\begin{equation}\label{eq:r and U/U0}
  dr = \frac{R^2 \sqrt{f^{-1}_y(y)\, h_y(y)}}{U_0\, y^2 \sqrt{y^4 - 1}}dy\,\, \Rightarrow\,\, r = \frac{R^2}{U_0}\int^{U/U_0}_{1} \frac{\sqrt{f^{-1}_y(y)\, h_y(y)}}{y^2 \sqrt{y^4 - 1}}dy.
\end{equation}
Hence, the radius $L$ of the Wilson loop obeys
\begin{equation}\label{eq:circular1loopl}
L = \frac{R^2}{U_0} \int^{\infty}_{1}\frac{\sqrt{f^{-1}_y(y)\, h_y(y)}}{y^2\sqrt{y^4-1}}dy\,.
\end{equation}
We can expand the integrand for small $U_T/U_0$ and large $C$ and then evaluate the integral analytically:
\begin{equation}\label{eq:L and U0}
L = \frac{R^2}{U_0}\left[\frac{\sqrt{\pi}\, \Gamma \left(\frac{3}{4}\right)}{\Gamma \left(\frac{1}{4}\right)} + \frac{R^4\, \Gamma \left(\frac{5}{4}\right)}{768\, \pi^{7/2}\, C^2 U_0^2\, \Gamma \left(\frac{7}{4}\right)} + \mathcal{O} \left(C^{-3}, \frac{U_T}{U_0}\right) \right]\, .
\end{equation}

On the other hand, $S_{\text{NG}}$ can be computed from \eqref{eq:circular1loope}. After some derivations and regularizations (see supplemental materials), we obtain the final result:
\begin{align}
S_{\text{NG}} & = \frac{\pi  R^2}{12} \left(- \, {_3F_2} \left(\frac{1}{2},\frac{1}{2},\frac{3}{4};1,\frac{7}{4};1\right) - \frac{24 \pi^2}{\Gamma \left(\frac{1}{4}\right)^4}\right) \nonumber\\
{} & \quad + \frac{B R^2\, \Gamma \left(\frac{1}{4}\right)^4}{2\pi^3} \frac{L^2}{C^2} + \mathcal{O} \left(C^{-3},\, U_T / U_0 \right)\, ,\label{eq:Circular Wilson Loop S_NG}
\end{align}
where
\begin{equation}
  B = \frac{1}{2304\, \pi^4} \Big[\pi -4 + 4\log (2)\Big]\, .
\end{equation}
This result of $S_{\text{NG}}$ implies that $\textrm{log}\, \langle W \rangle$ obeys the area law.

\begin{figure}[htb!]
\begin{center}
\includegraphics[width=7cm, angle=0]{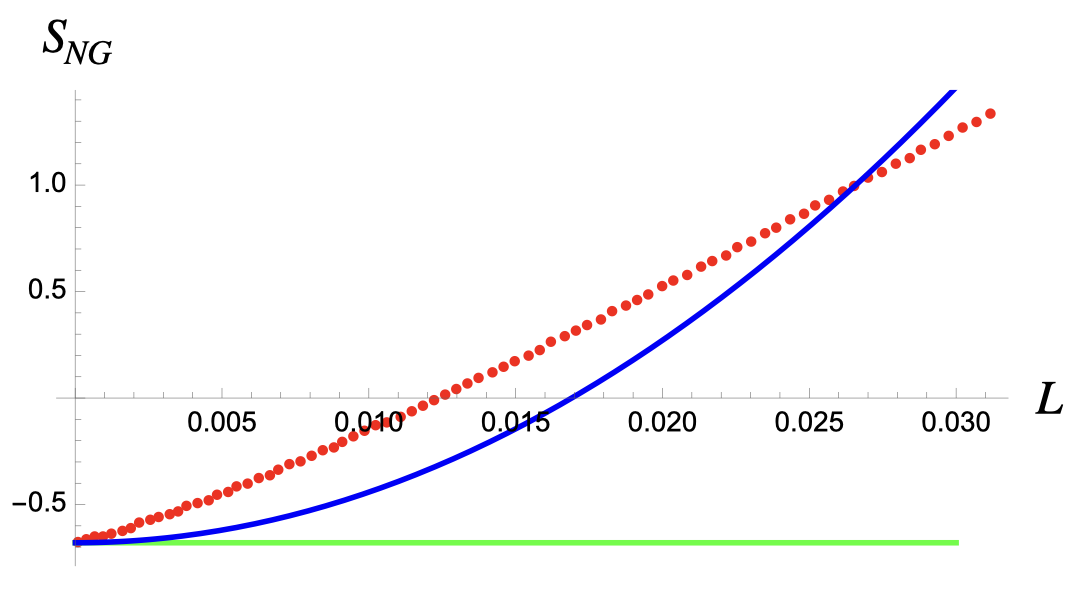}
\caption{The numerical result of $S_{\text{NG}}$ from the circular spatial Wilson loop with the parameters $R=1$, $Q=\sqrt{2}$, and $C=0.0001$}\label{fig:circular}
\end{center}
\end{figure}

We can also evaluate $S_{\text{NG}}$ numerically. The result is shown in Fig.~\ref{fig:circular}, where the analytic expression (blue curve) and the numerical result (red dots) clearly show area-law behavior for large $L$, significantly differing from the constant result for a conformal theory (green line) \cite{Berenstein:1998ij, Nian:2009mw}.

\textit{Discussion.---} The approach of introducing quantum gravity corrections in AdS/CFT correspondence already has many applications \cite{Lam:2018pvp, Iliesiu:2020qvm, Liu:2024gxr, Brown:2024ajk}. In this paper, we introduced the quantum gravity fluctuations in the near-horizon region of an extremal AdS$_5$ Reissner-Nordstr\"om black brane and computed the Wilson loop's quantum expectation value. We found the area law of $\textrm{log} \langle W \rangle$ and the linear term in the quark-antiquark potential, indicating the confinement in the boundary SYM theory. The deep insight we gained is that, from the viewpoint of the AdS/CFT correspondence, the quantum gravity fluctuations in the near-horizon region are in charge for the boundary large-$N$ SYM theory's confinement. A natural extension of the current work should provide a holographic interpretation for the SYM's mass gap \cite{MassGap}.

Although our results are derived within the framework of SYM theory, the quark-antiquark potential obtained in Eq.~\eqref{eq:Rectangular Wilson Loop Potential} can nonetheless be compared with the Cornell potential \cite{PhysRevLett.34.369, PhysRevD.17.3090, Brambilla:1999ja} to fix the parameters:
\be
  C \approx 0.09986\, \textrm{fm}\, ,\quad R = \frac{L_{\text{AdS}}}{\sqrt{\alpha'}} \approx 1.50863\, ,
\ee
which are reasonable values in nuclear phenomenology.

For an extremal RN black brane, the mass equals the charge and is proportional to the chemical potential. As the chemical potential increases, the black brane grows larger, and eventually, its horizon intersects the minimal surface anchored to the Wilson loop, which signals deconfinement, in full agreement with the SYM phase diagram (Fig.~\ref{fig:PhaseDiagram}). This behavior indicates a quantum-gravity-induced phase transition, which warrants further investigation. Conversely, in the case of a very small extremal RN black hole, the system approaches the origin of the phase diagram, but with a small deformation characterized by a nonzero chemical potential. This subtle deformation plays a crucial role in the emergence of confinement.

\section*{acknowledgments}

We would like to thank Mei Huang, Elias Kiritsis, Li Li, Jing Liu, Yu Tian, Gang Yang, and Hongbao Zhang for their helpful discussions. This work is supported in part by the NSFC under grants No.~12375067 and No.~12147103. J.N. would like to thank Lanzhou University and Northwest University for their warm hospitality during the final stage of this work.

X.-L. Liu and C.-Y. Yue contributed equally to this work.

\bibliography{QuWilsonLoop}




\appendix

\section{Quantum-Corrected AdS$_2$ Metric in Euclidean JT Gravity}

In this appendix, we generalize the formalism introduced in \cite{Blommaert:2019hjr} from the Lorentzian AdS$_2$ spacetime to the Euclidean one.

We begin with the Lorentzian AdS$_{2}$ metric in Poincar\'e coordinates:
\begin{equation}\label{eq:App AdS2 metric}
    ds^2 = \frac{1}{\widetilde{Z}^2} (d\widetilde{Z}\,^2 - dT^2)\,.
\end{equation}
By choosing $Z = i \widetilde{Z}$, we could get the Euclidean version of metric as
\begin{equation}
    ds^2 = \frac{1}{Z^2} (dZ^2 + dT^2)\, .
\end{equation}
For this bulk metric, a boundary observer lives in a wiggly curve $\widetilde{Z}(t) = \epsilon \dot{f}(t)$, where $f(t)$ is the reparametrization of boundary time $t$ \cite{Maldacena:2016upp}. It is should be noted that $f$ is the only degree of freedom in the theory.

The boundary observer can associate two boundary times with each bulk point by sending and receiving a lightray \cite{Blommaert:2019hjr}. More precisely, the observer sends a lightray at the boundary time $t_1 = v = t - \tilde{z}$, and then receive the lightray reflected by the bulk point at the boundary time $t_2 = u = t + \tilde{z}$.

In summary, the boundary observer's lightcone coordinates are defined as
\be
  v = t - \tilde{z}\, ,\quad u = t + \tilde{z}\, ,
\ee
while the bulk lightcone coordinates are defined as
\be
  V = T - \widetilde{Z}\, ,\quad U = T + \widetilde{Z}\, .
\ee
When there are no off-shell fluctuations, the bulk and the boundary lightcone coordinates coincide.

The process of relating a bulk point with two boundary times by a lightray is
\begin{equation}
    \begin{aligned}
        &\text{Send the light at:} & \quad v = t_1\, , & \quad T_1 = f(t_1)\, ;\\
        &\text{Receive the light at :} & \quad u = t_2\, , & \quad T_2 = f(t_2)\, .
    \end{aligned}
\end{equation}
With the identifications $V=T_{1}$ and $U=T_{2}$, we have
\begin{equation}
    U=f(u)\, ,\quad V=f(v)\, .
\end{equation}
Therefore, in the boundary observer's lightcone coordinates, we can rewrite the bulk AdS$_2$ metric \eqref{eq:App AdS2 metric} in boundary reparameterization modes $f$ as
\begin{equation}
    ds^2 = -\frac{f'(u)f'(v)}{(f(u)-f(v))^2}\, du\, dv\, .
\end{equation}

The line element can be derived from the AdS$_2$ geodesic distance, i.e., the isometric invariant distance, as
\begin{equation}\label{eq:App isometric invariant distance}
    d(P,Q)=\ln \Bigg|\frac{(U-V')(U'-V)}{(U-V)(U'-V')} \Bigg|\, .
\end{equation}
Quantum averaging the off-shell modes $f$, we obtain $\langle d_{f}(p,q)\rangle$ as the the exact bulk-bulk propagator
\begin{equation}
    \langle d_{f}(p,q)\rangle = \langle G_{bb}(u,u',v,v')\rangle = \int_{u'}^{u} dt \int_{v'}^{v} dt' \langle G_{\partial\partial}(t-t')\rangle\, ,
\end{equation}
where $\langle G_{\partial\partial}(t-t')\rangle$ is the zero-temperature exact boundary-boundary propagator in Lorentzian time given by
\begin{equation}
    \langle G_{\partial \partial}(t-t')\rangle = \frac{1}{2C} \int_{0}^{\infty} d\omega \sinh(2\pi\sqrt{2C\omega})\, e^{-i\omega(t-t')}\, \Gamma(1\pm i\sqrt{2C\omega})^2\, .
\end{equation}
By differentiating the isometric invariant distance \eqref{eq:App isometric invariant distance}, we obtain the line element as
\begin{align}
  \langle d_{f}(p,p+dx)\rangle & = \Bigg\langle \ln \Bigg| 1 - \frac{f(u)-f(u+du)}{f(u)-f(v)} \frac{f(v)-f(v+dv)}{f(u)-f(v)} \Bigg| \Bigg\rangle \nonumber\\
  & = \Bigg\langle \frac{f'(u)f'(v)}{(f(u)-f(v))^2} \Bigg\rangle\, du\, dv\, .
\end{align}
Hence, the quantum-averaged Lorentzian AdS$_2$ metric can be obtained by computing the exact boundary-boundary propagator and setting $t\to u$, $t'\to v$:
\begin{equation}
   \langle ds^2(t,\tilde{z}) \rangle= -\langle G_{\partial \partial}(u-v)\rangle\, du\, dv = \langle G_{\partial \partial}(2\tilde{z}) \rangle\, (d\tilde{z}^2-dt^2)\, ,
\end{equation}
with
\begin{equation}
    \langle G_{\partial \partial}(2\tilde{z})\rangle = \frac{1}{2C}\int_{0}^{\infty} d\omega \sinh(2\pi\sqrt{2C\omega})\, e^{-i\omega\tilde{z}}\, \Gamma(1\pm i\sqrt{2C\omega})^2\,.
\end{equation}
To obtain the quantum-corrected Euclidean AdS$_2$ metric, we transform $\tilde{z}\to -iz$, which consequently leads to
\be
  \langle ds^2(t,z)\rangle = \langle G_{\partial \partial}(2z) \rangle\, (dt^2 + dz^2)\, ,
\ee
where $\langle G_{\partial \partial}(2z) \rangle$ can be determined up to an overall proportionality constant as
\be
  \langle G_{\partial \partial}(2z)\rangle = \frac{1}{2C} \int_{0}^{\infty}d\omega \sinh(2\pi\sqrt{2C\omega})\, e^{-\omega z}\, \Gamma(1\pm i\sqrt{2C\omega})^2\, .
\ee

\section{Quantum-Corrected AdS$_5$ Black Brane Metric}

This appendix will discuss how the AdS$_5$ quantum correction factor is introduced via the matching process.

The main process of adding quantum corrections to the AdS$_5$ metric consists of the following few steps:
\begin{enumerate}
    \item Separate the original AdS$_{5}$ black brane metric into a near-horizon region (NHR) and a far-away region (FAR).
 
    \item Since the NHR metric becomes AdS$_{2}\times \mathbb{T}^3$, by performing a dimensional reduction \cite{Iliesiu:2020qvm}, we obtain a JT gravity sector, which controls the low-temperature quantum fluctuation behavior of the black brane.

    \item By transforming into the NHR boundary coordinates \cite{Blommaert:2019hjr}, we can rewrite the NHR metric in terms of the Schwarzian mode, which is the zero mode that emerge in the NHR boundary when doing the JT path integral.

    \item Finally, by integrating all Schwarzian modes in the NHR metric, we obtain the quantum-corrected NHR metic. Since the quantum-corrected metric should be continuous and smooth in the overlap region of NHR and FAR, we obtain the quantum-corrected FAR metric by matching the asymptotic behaviors of both regions.
\end{enumerate}

The near-extremal RN AdS$_5$ black brane has the metric in the global coordinates:
\begin{equation}\label{eq:App RN BH metric 1}
  ds^2 = -\frac{u^2}{L_{\text{AdS}}^2} f(u)\, dt^2 + \frac{L_{\text{AdS}}^2}{u^2} \frac{du^2}{f(u)} + \frac{u^2}{L_{\text{AdS}}^2} d\vec{x}\,^2\, ,
\end{equation}
with
\begin{equation}
  f(u) = 1 - (1 + Q^2)\frac{u_T^4}{u^4} + Q^2\frac{u_T^6}{u^6}\, ,\quad T_h = \frac{(2-Q^2) u_T}{2\pi L_{\text{AdS}}^2}\, .
\end{equation}

Now, let us take the extremal limit $Q^2 = 2$ and define a small parameter
\be
  \lambda \equiv \frac{u-u_T}{u_T}
\ee
to probe the near-horizon region. Taking the $\lambda\to0$ limit and keeping only the leading order, the metric becomes
\begin{equation}
ds^2 = -\frac{12 (u-u_T)^2}{L^2_{\text{AdS}}} dt^2 + \frac{L^2_{\text{AdS}}}{12 (u-u_T)^2} du^2 + \frac{u_T^2}{L^2_{\text{AdS}}} d\vec{x}\,^2\, .
\end{equation}
In order to introduce quantum gravity corrections, we need to adopt a further change of coordinates:
\begin{equation}
  u = u_T+\frac{L_{\text{AdS}}^2}{12\zeta}
\end{equation}
to bring the near-horizon metric into a form with an explicit AdS$_2$ part. Consequently, the metric becomes
\begin{equation}
ds^2 = -\frac{L^2_{\text{AdS}}}{12\zeta^2} dt^2 + \frac{L^2_{\text{AdS}}}{12\zeta^2} d\zeta^2 + \frac{u_T^2}{L^2_{\text{AdS}}} d\vec{x}\,^2\, ,
\end{equation}
where the AdS$_2$ part is exactly the same as in \cite{Blommaert:2019hjr}:
\begin{equation}
ds^2 = -\frac{L_{\text{AdS}}^2}{12\zeta^2}dt^2 + \frac{L_{\text{AdS}}^2}{12\zeta^2}d\zeta^2 = -\frac{L_2^2}{\zeta^2}dt^2 +\frac{L_2^2}{\zeta^2}d\zeta^2\, .
\end{equation}
We have introduced an AdS$_2$ radius $L_2$, which is related to the AdS$_5$ radius $L_{\text{AdS}}$ via
\begin{equation}
L_2 \equiv \frac{L_{\text{AdS}}}{\sqrt{12}}\, .
\end{equation}

In this near-horizon AdS$_2$, we perform a Wick rotation and introduce the quantum correction factor $h(\zeta)$ as in \cite{Blommaert:2019hjr}:
\begin{equation}
h(\zeta) = 1 + \frac{3}{2 \pi ^4 }\left(\frac{\zeta}{C}\right)^2 - \frac{15 }{2 \pi ^6 }\left(\frac{\zeta}{C}\right)^3 + \frac{315}{8 \pi ^8 }\left(\frac{\zeta}{C}\right)^4 +\cdots\, .
\end{equation}
Next, we can apply an inverse coordinate transformation:
\begin{equation}
\zeta = \frac{L^2_{\text{AdS}}}{12(u-u_T)}\, ,
\end{equation}
to lift the near-horizon AdS$_2$'s quantum correction factor back to the original AdS$_5$ black brane coordinates. Then, the quantum correction factor $h$ in the original coordinate $u$ is given by
\begin{align}
h(u) & = 1 + \frac{L_{\text{AdS}}^4}{96\, \pi ^4 C^2 (u-u_T)^2} -\frac{5 L_{\text{AdS}}^6}{1152\, \pi ^6C^3  (u-u_T)^3} \nonumber\\
{} & \quad + \frac{35 L_{\text{AdS}}^8}{18432\, \pi ^8 C^4 (u-u_T)^4}+\cdots\, .\label{eq:quantum correction factor in App}
\end{align}
For later convenience, we define another coordinate $U$ via
\begin{equation}
U \equiv \frac{u}{\alpha'}\, ,\quad U_T \equiv \frac{u_T}{\alpha'}\, ,\quad\text{with } \alpha' = \frac{L_{\text{AdS}}^2}{R^2}\, ,
\end{equation}
where $\sqrt{\alpha'}$ denotes the effective string length. Finally, the quantum-corrected near-horizon AdS$_5$ metric becomes
\begin{equation}\label{eq:quantum corrected AdS2 metric in App}
\begin{aligned}
    \langle ds^2_{NHR}\rangle & = \alpha' \Bigg[-h(U)\frac{12(U-U_T)^2}{R^2}dt^2+h(U)\frac{R^2}{12(U-U_T)^2} dU^2 \\
    {} & \qquad +\frac{U_T^2}{R^2} d\vec{x}\,^2\Bigg]\, ,
\end{aligned}
\end{equation}
and the quantum correction factor is
\begin{align}
h(U) & = 1 + \frac{R^4}{96\, \pi ^4 C^2 (U-U_T)^2} -\frac{5 R^6}{1152\, \pi ^6C^3  (U-U_T)^3} \nonumber\\
{} & \quad + \frac{35 R^8}{18432\, \pi ^8 C^4 (U-U_T)^4}+\cdots\, ,\label{eq:quantum correction factor in App}
\end{align}
where the dots denote higher-order subleading terms.

In principle, \eqref{eq:quantum corrected AdS2 metric in App} is the quantum-corrected metric in the near-horizon region, and the quantum correction factor has the asymptotic expression \eqref{eq:quantum correction factor in App}, which can only be used in the NHR. The full quantum-corrected AdS$_5$ metric should be obtained by requiring the continuous and smooth boundary conditions in the overlap region of the NHR and the FAR, i.e.,
\begin{equation}\label{eq:junction condition}
    \langle ds^2_{\text{NHR}} \rangle \big|_{\text{overlap}}=\langle ds^2_{\text{FAR}}\rangle \big|_{\text{overlap}}\, .
\end{equation}
This condition fixes the quantum-corrected metric near the AdS$_5$ boundary to be
\begin{equation}\label{eq:quantum corrected AdS5 metric in App}
  ds^2 = \alpha'\left[\frac{U^2}{R^2} \left(f(U)\, h(U)\, dt^2+d\vec{x}\,^2\right) + R^2f^{-1}(U)\, h(U)\frac{dU^2}{U^2} \right] \, ,
\end{equation}
which is the most relevant to the computations in this paper. Since the metric \eqref{eq:quantum corrected AdS2 metric in App} with the quantum correction factor \eqref{eq:quantum correction factor in App} can provide the asymptotic behavior near the NHR boundary by taking $(U-U_T)/U_T \gg 1$. Then, we can match it with the asymptotic behavior of the FAR in the overlap region using the condition \eqref{eq:junction condition}. We see that the $U\to \infty$ asymptotics of \eqref{eq:quantum corrected AdS2 metric in App} are consistent with \eqref{eq:quantum corrected AdS5 metric in App}. Hence, we take \eqref{eq:quantum corrected AdS5 metric in App} as the quantum-corrected AdS$_5$ metric in the main text.

\section{Subleading Quantum Correction Factor}

In this appendix, we explain why the quantum correction factor only appears in the $(t, u)$-directions of the AdS$_5$ metric,  or in other words, the spatial dimensions $d\vec{x}\,^2$ do not receive quantum corrections at the leading order.

Let us still begin with the AdS$_5$ RN black brane metric in the global coordinates:
\begin{equation}
    ds^2 = -\frac{u^2}{L_{\text{AdS}}^2} f(u)\, dt^2 + \frac{L_{\text{AdS}}^2}{u^2}\frac{du^2}{f(u)} + \frac{u^2}{L_{\text{AdS}}^2}d\vec{x}\,^{2},
\end{equation}
with
\begin{equation}
    f(u) = 1 - (1 + Q^2)\frac{u_{T}^4}{u^4} + Q^2 \frac{u_{T}^6}{u^6},\quad T = \frac{(2 - Q^2) u_{T}}{2\pi L_{\text{AdS}}^2},
\end{equation}
where $u_{T}$ denotes the horizon location, and $\vec{x}$ includes the coordinates $(r,x,y)$.

Besides the $(t, u)$-coordinates, we also expand the $\vec{x}$-directions of the metric near $(u - u_{T})/u_{T}$. Then, we obtain the near-horizon metric as
\begin{align}
  ds^2 & = -\frac{(u-u_{T})^2}{L_{2}^2} dt^2 + \frac{L_{2}^2}{(u - u_{T})^2} du^2\nonumber\\
  {} & \quad + \left(\frac{u_{T}^2}{12L_{2}^2} + \frac{2 (u - u_{T}) u_{T}}{12L_{2}^2} + \frac{(u - u_{T})^2}{12L_{2}^2}\right) d\vec{x}\,^2\, ,
\end{align}
where $L_{2} \equiv L_{\text{AdS}}/\sqrt{12}$ is the AdS$_2$ radius in the near-horizon region. Note that in the near-horizon limit, the second line can be simplified into a leading constant term, $u_{T}^2/(12 L_{2}^2)$, since the other two terms are subleading compared to this term.

By defining a new coordinate transformation, $\zeta \equiv \frac{L_{2}^2}{u - u_T}$, we rewrite the NHR metric as
\begin{equation}
    ds^2 = -\frac{L_{2}^2}{\zeta^2}(-dt^2+d\zeta^2) + \frac{u_T^2}{12L_{2}^2} \left(1 + \frac{2L_{2}^2}{u_{T}\zeta} + \frac{L_{2}^4}{u_{T}^2\zeta^2}\right) d\vec{x}\,^2\, ,
\end{equation}
from which we see that, by only keeping the leading-order terms, the metric in the NHR could be factorized into AdS$_{2}\times \mathbb{T}^3$.

Now, we consider quantum corrections to this metric. The quantum correction to the AdS$_{2}$ part can be introduced similarly as in \cite{Blommaert:2019hjr}. We first define the light-cone coordinates $x_\pm$ via
\be
  t = \frac{x_{+} + x_{-}}{2}\, ,\quad \zeta = \frac{x_{+} - x_{-}}{2}\, .
\ee
Consequently, the metric in the light-cone coordinates becomes
\begin{align}
    ds^2_{\text{\text{AdS}}_{2}\times \mathbb{T}^3} & = \frac{4 L_{2}^2}{(x_{+} - x_{-})^2}\, dx_{+} dx_{-} \nonumber\\
    {} & \quad + \frac{u_T^2}{12 L_{2}^2} \left(1 + \frac{4 L_{2}^2}{u_{T} (x_{+} - x_{-})} + \frac{4L_{2}^4}{u_{T}^2 (x_{+} - x_{-})^2}\right) d\vec{x}\,^2\, .
\end{align}
These light-cone coordinates for a boundary observer can be lifted to Schwarzian modes when the quantum fluctuations of the NHR boundary are turned on. Thus, we have
\begin{equation}
    ds^2_{\text{\text{AdS}}_{2}} = \frac{4L_{2}^2\, F'(x_{+})\, F'(x_{-})}{(F(x_{+}) - F(x_{-}))^2}\, dx_{+}\, dx_{-}\, .
\end{equation}
We can define a Schwarzian bilocal operator:
\begin{equation}
    O(x_{+}, x_{-}) \equiv \frac{F'(x_{+})\, F'(x_{-})}{\bigl(F(x_{+}) - F(x_{-})\bigr)^2}\, ,
\end{equation}
whose expectation value can be calculated in both the exact and the perturbative methods \cite{Mertens:2017mtv}. Therefore, the quantum-corrected AdS$_{2}$ metric can be expressed as
\begin{equation}
    \langle ds_{\text{AdS}_{2}}^2\rangle = 4 L_2^2\, \langle O(x_{+},x_{-})\rangle\, dx_{+} dx_{-}.
\end{equation}
By performing an inverse transformation back to the AdS$_{2}$ Poincar\'e coordinates, we rewrite the quantum-corrected metric in the form of the classical metric with a quantum correction factor $h(\zeta)$, i.e.,
\begin{equation}
    \langle ds_{\text{AdS}_{2}}^2\rangle = h(\zeta)\, \frac{L_2^2}{\zeta^2}(-dt^2 + d\zeta^2)\, .
\end{equation}

For the $\mathbb{T}^3$-part of the near-horizon metric, the subleading terms also contain light-cone coordinates, which receive quantum fluctuations when lifted to the Schwarzian mode. Hence, in principle, we should also consider the quantum correction to the $\mathbb{T}^3$-part of the near-horizon metric. However, we will show that this quantum correction is more subleading and can be ignored. The $\mathbb{T}^3$-part of the near-horizon metric with subleading terms is
\begin{equation}
    ds^2_{\mathbb{T}^3} = \frac{u_T^2}{12L_{2}^2} \left(1 + \frac{4L_{2}^2}{u_{T}(x_{+} - x_{-})} + \frac{4L_{2}^4}{u_{T}^2 (x_{+} - x_{-})^2}\right)\, d\vec{x}\,^2\, ,
\end{equation}
with $u_T / L_{2}\gg 1$. After lifting the light-cone coordinates, $x_\pm$, to the Schwarzian mode, we can derive the quantum correction from the Schwarzian mode's correlation functions, which can be evaluated perturbatively by using $F(t) = t + \epsilon(t)$ and expanding in small $\epsilon(t)$. More explicitly,
    \begin{align}
       \bigg\langle \frac{1}{F(x_{+}) - F(x_{-})}\bigg\rangle & = \frac{1}{x_{+} - x_{-}} + \frac{\langle(\epsilon(x_{+}) - \epsilon(x_{-}))^2\rangle}{(x_{+} - x_{-})^3} + \mathcal{O}(\epsilon^3)\, ,\\
       \bigg\langle \frac{1}{(F(x_{+}) - F(x_{-}))^2}\bigg\rangle & = \frac{1}{(x_{+} - x_{-})^2} + \frac{3\langle (\epsilon(x_{+}) - \epsilon(x_-)^2)\rangle}{(x_{+} - x_{-})^2} + \mathcal{O}(\epsilon^3)\, .
    \end{align}
Therefore, the quantum-corrected $\mathbb{T}^3$-part of the metric with subleading terms is
\begin{equation}
\begin{aligned}
    \langle ds^2_{\mathbb{T}^3}\rangle & = \frac{u_T^2}{12L_{2}^2} \Bigg[1 + \frac{4L_{2}^2}{u_{T} (x_{+} - x_{-})} + \frac{4L_{2}^2}{u_{T}} \left(\frac{\langle(\epsilon(x_{+}) - \epsilon(x_{-}))^2\rangle}{(x_{+} - x_{-})^3}\right)\\
    & + \frac{4L_{2}^4}{u_{T}^2(x_{+} - x_{-})^2} + \frac{4L^2_{2}}{u_{T}^2} \left(\frac{3\langle (\epsilon(x_{+}) - \epsilon(x_-))^2\rangle}{(x_{+} - x_{-})^2}\right) + \mathcal{O}(\epsilon^3)\Bigg] d\vec{x}\,^2.
\end{aligned}  
\end{equation}
As we can see, the first quantum correction appears at the order $\mathcal{O} (L_{2}^2 \epsilon^2 / u_{T})$ or $\mathcal{O} (L_{2}^2 \epsilon^2 / u_{T}^2)$.

On the other hand, the quantum correction to the AdS$_{2}$-part is carried by the bilocal operator, which can also be evaluated in a perturbative way as follows:
\begin{equation}
\begin{aligned}
{} & \left\langle \frac{F'(x_{+})F'(x_{-})}{(F(x_{+})-F(x_{-}))^2}\right\rangle \nonumber\\
=\, & \frac{1}{(x_{+}-x_{-})^2}+\frac{3\langle(\epsilon(x_{+})-\epsilon(x_{-}))^2\rangle}{(x_{+}-x_{-})^4}-\frac{2\langle (\epsilon(x_{+})-\epsilon(x_{-}))\epsilon'(x_{+})\rangle}{(x_{+}-x_{-})^3} \nonumber\\
{} & -\frac{2\langle (\epsilon(x_{+})-\epsilon(x_{-}))\epsilon'(x_{-})\rangle}{(x_{+}-x_{-})^3}+\frac{\langle\epsilon'(x_{+})\epsilon'(x_{-})\rangle}{(x_{+}-x_{-})^2}+\mathcal{O}(\epsilon^3).
\end{aligned}
\end{equation}
Hence, the leading quantum correction to the AdS$_{2}$-part is of order $\mathcal{O}(\epsilon^2)$, while the $\mathbb{T}^3$-part is of order $\mathcal{O}(L_{2}\epsilon^2 / u_{T})$, which indicates that the Schwarzian fluctuation's contribution to the $\mathbb{T}^3$-part is further suppressed in the NHR (due to $u_T / L_2 \gg 1$) and can be ignored compared to the leading-order corrections. We should note that this argument is valid for the NHR only, and the AdS$_{2}$ Poincar\'e coordinates $(t, \zeta)$ cannot cover the full AdS$_{5}$, taking $\zeta\to 0$ actually corresponding to the AdS$_{2}$ boundary. Therefore, we finally get a quantum-corrected NHR metric as
\begin{equation}
    \langle ds^2_{\text{NHR}}\rangle=h(\zeta)\frac{L^2_{2}}{\zeta^2}(-dt^2+d\zeta^2)+\frac{u_T^2}{12L_{2}^2}d\vec{x}\,^2.
\end{equation}
By imposing the continuous and smooth conditions in the overlap region, we match the NHR and the FAR metrics as follows:
\begin{equation}
    \langle ds^2_{\text{NHR}} \rangle \big|_{\text{overlap}}=\langle ds^2_{\text{FAR}}\rangle \big|_{\text{overlap}}\, .
\end{equation}
Thus, we can glue on the quantum correction factor to the FAR metric to obtain the quantum-corrected RN AdS$_{5}$ black brane metric:
\begin{equation}
    \langle ds^2 \rangle = - h(u)\frac{u^2}{L^2}f(u)dt^2 + h(u)\frac{L^2}{u^2}\frac{du^2}{f(u)} + \frac{u^2}{L^2} d\vec{x}\,^{2}\, .
\end{equation}
When we evaluate the quantum correction factor $h(u)$ in FAR, it corresponds to the near-boundary expansion of $h(u)$.

\section{Details for the Circular Spatial Wilson Loop}

For the circular spatial Wilson loop, the NG action for the minimal surface anchored to the Wilson loop can be expressed as an integral \eqref{eq:circular1loope}. In this appendix, we evaluate it and regularize the result explicitly. The regularization schemes we are using include
\begin{itemize}
\item[a)] Evaluate the divergent integral with a finite cutoff and introduce a corresponding counter-term to cancel the divergence;

\item[b)] Evaluate the divergent integral with a finite cutoff and only keep the finite terms.
\end{itemize}

We start with the expression in \eqref{eq:circular1loope}:
\begin{align}
S_{NG} & =\int_0^L r dr\sqrt{(\partial_rU)^2 f^{-1}(U)\, h(U) + U^4/R^4} \nonumber\\
{} & =\int_1^\infty r\frac{U^4}{U^2_0R^2}\frac{U_0 R^2\sqrt{f^{-1}_y(y)\, h_y(y)}}{U^2 \sqrt{\frac{U^4}{U^4_0}-1}} dy \nonumber\\
{} & = \frac{1}{U_0}\int_1^\infty r\frac{U^2\sqrt{f^{-1}_y(y)\, h_y(y)}}{\sqrt{\frac{U^4}{U^4_0}-1}} dy = U_0\int_1^\infty r\frac{y^2\sqrt{f^{-1}_y(y)\, h_y(y)}}{\sqrt{y^4-1}}dy\nonumber\\
{} & = U_0\int^{\infty}_1 dy\frac{y^2\sqrt{f^{-1}_y(y)\, h_y(y)}}{\sqrt{y^4-1}} \left(\frac{R^2}{U_0}\int^{U/U_0}_{1} \frac{\sqrt{f^{-1}_x(x)\, h_x(x)}}{x^2 \sqrt{x^4-1}} dx\right)\nonumber\\
{} & = R^2\int^{\infty}_1 dy\frac{y^2\sqrt{f^{-1}_y(y)\, h_y(y)}}{\sqrt{y^4-1}} \left(\int^{y}_{1}\frac{\sqrt{f^{-1}_x(x)\, h_x(x)}}{x^2\sqrt{x^4-1}}dx\right)\, ,\label{eq:circular WL S_NG intermediate}
\end{align}
where we have used Eq.~\eqref{eq:r and U/U0}.

To evaluate the integral inside the brackets in the last expression of \eqref{eq:circular WL S_NG intermediate}, we first expand the integrand: 
\begin{align}
\frac{\sqrt{f^{-1}_x(x)\, h_x(x)}}{x^2\sqrt{x^4-1}} & = \frac{1}{x^2 \sqrt{x^4-1}} + \frac{R^4}{192\, \pi^4 C^2 U_0^2 x^4 \sqrt{x^4-1}}\nonumber\\
{} & \quad + \mathcal{O} \left(C^{-3}, U_T/U_0\right)\,.
\end{align}
Hence, the integral consists of two parts in this order. The first part is the classical result:
\begin{equation}
\frac{\sqrt{\pi}\, \Gamma \left(\frac{3}{4}\right)}{\Gamma \left(\frac{1}{4}\right)} - \frac{\, _2F_1\left(\frac{1}{2},\frac{3}{4};\frac{7}{4};\frac{1}{y^4}\right)}{3 y^3}\, ,
\end{equation}
while the second part is the quantum correction result:
\begin{align}
\frac{R^4}{2304\, \pi ^4 C^2 U_0^2}\, \Bigg[ & 4 y^3 \, _2F_1\left(-\frac{3}{4},\frac{1}{2};\frac{1}{4};\frac{1}{y^4}\right)-\frac{4 \left(y^4-1\right)^{3/2}}{y^3}\nonumber\\
{} & + \frac{3 \sqrt{\pi}\, \Gamma \left(\frac{5}{4}\right)}{\Gamma \left(\frac{7}{4}\right)} \Bigg]\, .
\end{align}
Besides the integral inside the brackets, there is another term in the integrand of the last line of \eqref{eq:circular WL S_NG intermediate}. We can also expand it in a similar way:
\begin{align}
\frac{y^2\sqrt{f^{-1}_y(y)\, h_y(y)}}{\sqrt{y^4-1}} & = \frac{y^2}{\sqrt{y^4-1}} + \frac{R^4}{192\, \pi^4 C^2 U_0^2 \sqrt{y^4-1}}\nonumber\\
{} & \quad + \mathcal{O} \left(C^{-3},U_T/U_0\right)\,.
\end{align}
Combining these expressions, we evaluate the last line of \eqref{eq:circular WL S_NG intermediate}, which can also be written into two parts. For the classical part, we regularize it as follows:
\begin{align}
&\int_1^{\infty } \left(\frac{y^2}{\sqrt{y^4-1}} - 1\right) \left(\frac{\sqrt{\pi}\, \Gamma \left(\frac{3}{4}\right)}{\Gamma \left(\frac{1}{4}\right)}-\frac{\, _2F_1\left(\frac{1}{2},\frac{3}{4};\frac{7}{4};\frac{1}{y^4}\right)}{3 y^3}\right) \, dy \nonumber\\
{} & - \left[\frac{\pi}{4} + \sqrt{\pi} \left(\frac{\Gamma \left(\frac{3}{4}\right)}{\Gamma \left(\frac{1}{4}\right)}-\frac{\Gamma \left(\frac{7}{4}\right)}{3\, \Gamma \left(\frac{5}{4}\right)}\right)\right]\, ,
\end{align}
and the result is
\begin{equation}
-\frac{1}{12} \pi\, {_3F_2} \left(\frac{1}{2},\frac{1}{2},\frac{3}{4};1,\frac{7}{4};1\right)-\frac{\sqrt{2} \pi ^2 \Gamma \left(\frac{3}{4}\right)}{\Gamma \left(\frac{1}{4}\right)^3}\, .
\end{equation}

For the quantum correction part, keeping terms up to the order $1/C^2$, the integral is 
\begin{align}
{} & \frac{R^4}{2304\, \pi ^4 C^2 U_0^2} \int^{\infty}_1 dy\frac{1}{\sqrt{y^4-1}} \Bigg[\frac{12 \sqrt{\pi}\, \Gamma \left(\frac{3}{4}\right)}{\Gamma \left(\frac{1}{4}\right)} - \frac{4 \, _2F_1\left(\frac{1}{2},\frac{3}{4};\frac{7}{4};\frac{1}{y^4}\right)}{y^3} \nonumber\\
{} & + 4 y^5 \, _2F_1\left(-\frac{3}{4},\frac{1}{2};\frac{1}{4};\frac{1}{y^4}\right) -\frac{4 \left(y^4-1\right)^{3/2}}{y} + \frac{3 \sqrt{\pi}\, y^2\, \Gamma \left(\frac{5}{4}\right)}{\Gamma \left(\frac{7}{4}\right)} \Bigg]\, .\label{eq:Integral for Circular Wilson Loop}
\end{align}
Schematically, we can denote this integral as 
\begin{equation}
B\frac{R^4}{C^2\, U_0^2}\, ,
\end{equation}
where $B$ is a divergent constant. We should regularize this expression and extract the finite piece as the final result. There are five terms in the integrand of \eqref{eq:Integral for Circular Wilson Loop}. Let us integrate them and regularize the result term by term.
\begin{enumerate}

    \item[(1)] The first term in the integrand of \eqref{eq:Integral for Circular Wilson Loop} is
    \begin{equation}
        \frac{12\sqrt{\pi}\, \Gamma(\frac{3}{4})}{\Gamma(\frac{1}{4})\sqrt{y^4-1}}\, .
    \end{equation}
    A direct integration leads to
    \begin{equation}
        \int_{1}^{\infty}dy\frac{12\sqrt{\pi}\, \Gamma(\frac{3}{4})}{\Gamma(\frac{1}{4})\sqrt{y^4-1}} = 3\pi.
    \end{equation}
    This result does not need to be regularized.

\item[(2)] The second term in the integrand of \eqref{eq:Integral for Circular Wilson Loop} is
\begin{equation}
    -\frac{4\, {_{2}F_{1}(\frac{1}{2}, \frac{3}{4}; \frac{7}{4}; \frac{1}{y^4})}}{y^3\sqrt{y^4-1}}\, .
\end{equation}
We make a change of coordinate, $z = 1 / y^4$, then
\begin{align}
  {} & \int^{\infty}_1 dy \left[-\frac{4\; {_{2}F_{1}} (\frac{1}{2}, \frac{3}{4}; \frac{7}{4}; \frac{1}{y^4})}{y^3\sqrt{y^4-1}}\right] \nonumber\\
  =\, & \int^0_1 dz\,  \, _2F_1\left(1,\frac{5}{4};\frac{7}{4};z\right)\, \\
  =\, & \log (8)-\frac{3 \pi }{2}\, .
  \end{align}

This result does not need to be regularized.

    \item[(3)] The third term in the integrand of \eqref{eq:Integral for Circular Wilson Loop} is
    \begin{equation}
        \frac{4y^5\; {_{2}F_{1}}(-\frac{3}{4}, \frac{1}{2}; \frac{1}{4}; \frac{1}{y^4})}{\sqrt{y^4-1}}\, .
    \end{equation}
To obtain an analytical result, we first define $z = 1 / y^4$, then the integral becomes
 \begin{align}
   \int^{\infty}_1 dy\frac{4y^5\; {_{2}F_{1}}(-\frac{3}{4}, \frac{1}{2}; \frac{1}{4}; \frac{1}{y^4})}{\sqrt{y^4-1}} = \int^1_0 dz \frac{\, _2F_1\left(-\frac{1}{4},1;\frac{1}{4};z\right)}{z^2}\, .
 \end{align}
In order to regularize the result, we introduce a cutoff $\Lambda$ as follows:
 \begin{align}
    &\int^1_{\Lambda} dz \frac{\, _2F_1\left(-\frac{1}{4},1;\frac{1}{4};z\right)}{z^2}\nonumber\\
   =\, & -\frac{\Gamma \left(\frac{1}{4}\right) G_{3,3}^{2,2}\left(-\Lambda\left|
\begin{array}{c}
 0,\frac{5}{4},2 \\
 0,1,\frac{3}{4} \\
\end{array}
\right.\right)}{\Lambda \Gamma \left(-\frac{1}{4}\right)}-\frac{\Gamma \left(\frac{1}{4}\right) G_{3,3}^{2,2}\left(-1\left|
\begin{array}{c}
 0,\frac{5}{4},2 \\
 0,1,\frac{3}{4} \\
\end{array}
\right.\right)}{\Gamma \left(-\frac{1}{4}\right)}\nonumber\\
   =\, & \frac{1}{\Lambda} + \frac{2 \sqrt{2}\, \pi ^{3/2}}{\sqrt[4]{\Lambda}\, \Gamma \left(\frac{3}{4}\right)^2} + 4 \log (\Lambda) + \left(\frac{\pi }{2}-5+\log (2)\right)+ \mathcal{O}\left(\Lambda\right)\, ,
    \end{align}
where $G^{m,n}_{p,q} \biggl(z\, \Big|\begin{array}{c}
  \bf{a}\\\bf{b}
\end{array}\biggr)$ is the Meijer G-function, and in the second equality, we expand in $\Lambda\to 0$. Then, we obtain after the regularization:
 \begin{equation}
\int^{\infty}_1 dy\, \frac{4y^5\; {_{2}F_{1}}(-\frac{3}{4}, \frac{1}{2}; \frac{1}{4}; \frac{1}{y^4})}{\sqrt{y^4-1}} = \frac{ \pi }{2}-5+\log (2)\, .
\end{equation}

    \item[(4)] The fourth term in the integrand of \eqref{eq:Integral for Circular Wilson Loop} is
    \begin{equation}
        -\frac{4(y^4-1)^{3/2}}{y\sqrt{y^4 - 1}}\, .
    \end{equation}
A direct integration leads to
\begin{equation}
\int_{1}^{A} dy \left(-\frac{4(y^4-1)^{3/2}}{y\sqrt{y^4-1}}\right) = 1 - A^4 + 4\log A\, ,
\end{equation}
which diverges for $A\to\infty$. Thus, we can introduce counter-terms $-4y^3 + \frac{4}{y}$ in the integrand. Then, the regularized integral is finite:
\begin{equation}
\int_{1}^{\infty} dy\, \left(-\frac{4(y^4-1)^{3/2}}{y\sqrt{y^4-1}}+4y^3-\frac{4}{y}\right) + 1 = 1\, .
\end{equation}

    \item[(5)] The fifth term in the integrand of \eqref{eq:Integral for Circular Wilson Loop} is
    \begin{equation}
        \frac{3\sqrt{\pi}\, \Gamma(\frac{5}{4})}{\Gamma(\frac{7}{4})}\frac{y^2}{\sqrt{y^4-1}}.
    \end{equation}
    A direct integration leads to
    \begin{equation}
        \int_{1}^{A} dy \frac{y^2}{\sqrt{y^4-1}} = -\frac{i}{3} \left(\frac{\sqrt{\pi}\, \Gamma \left(\frac{7}{4}\right)}{\Gamma\, \left(\frac{5}{4}\right)}-A^3 \, _2F_1\left(\frac{1}{2},\frac{3}{4};\frac{7}{4};A^4\right)\right)\, .
    \end{equation}
Because of the following asymptotic behaviors:
\begin{align}
\lim_{A\to\infty}  \, _2F_1\left(\frac{1}{2},\frac{3}{4};\frac{7}{4};A^4\right) & = 0\, ,\\
\lim_{A\to\infty} A^2 \, _2F_1\left(\frac{1}{2},\frac{3}{4};\frac{7}{4};A^4\right) = -\frac{4 i\, \Gamma \left(\frac{7}{4}\right)}{\Gamma \left(\frac{3}{4}\right)} & = -3i\, ,\\
\lim_{A\to\infty} A^3 \, _2F_1\left(\frac{1}{2},\frac{3}{4};\frac{7}{4};A^4\right) & = (-i)\infty\, ,
\end{align}
we can conclude that the divergence of the integral is of order $\mathcal{O}(A)$ for $A \to \infty$. Hence, we only need to introduce a constant $1$ in the integrand as the counter-term. Consequently, the regularized integral becomes

\begin{equation}
\frac{3\sqrt{\pi}\, \Gamma(\frac{5}{4})}{\Gamma(\frac{7}{4})} \left[\int_{1}^{\infty} dy\, \left(\frac{y^2}{\sqrt{y^4-1}} - 1\right)-1\right] = -\pi\, .
\end{equation}

\end{enumerate}

In summary, we obtain the regularized result for the integral in \eqref{eq:Integral for Circular Wilson Loop}:
\begin{align}
{} & 3\pi + \log (8)-\frac{3 \pi }{2} + \frac{ \pi }{2}-5+\log (2) +1 -\pi \nonumber\\
=\, & \pi -4 + 4\log (2)\,.
\end{align}
Combining the classical and the quantum correction parts, the NG action is given by
\begin{align}
S_{NG} & = R^2 \Bigg[-\frac{1}{12} \pi  \, _3F_2\left(\frac{1}{2},\frac{1}{2},\frac{3}{4};1,\frac{7}{4};1\right)-\frac{\sqrt{2} \pi ^2 \Gamma \left(\frac{3}{4}\right)}{\Gamma \left(\frac{1}{4}\right)^3}\nonumber\\
{} & \qquad + B\frac{R^4}{U_0^2C^2} \Bigg]\, ,
\end{align}
where
\begin{equation}
B = \frac{1}{2304\, \pi^4} \Big[\pi -4 + 4\log (2)\Big]\, .
\end{equation}
Finally, taking into account the relation \eqref{eq:L and U0} between $L$ and $U_0$, we obtain
\begin{align}
S_{NG} & = \frac{1}{12} \pi  R^2 \left(-\, _3F_2\left(\frac{1}{2},\frac{1}{2},\frac{3}{4};1,\frac{7}{4};1\right)-\frac{12 \sqrt{2} \pi\, \Gamma \left(\frac{3}{4}\right)}{\Gamma \left(\frac{1}{4}\right)^3}\right)\nonumber\\ 
{} & \quad +\frac{B R^2 \Gamma \left(\frac{1}{4}\right)^2}{\pi  \Gamma \left(\frac{3}{4}\right)^2} \frac{L^2}{C^2} + \mathcal{O} \left(C^{-3}, \frac{U_T}{U_0}\right)\, .
\end{align}

\bibliographystyle{utphys}

\end{document}